\documentclass[12pt,a4paper]{article}
\usepackage{geometry}
\usepackage[english]{babel}
\usepackage{csquotes}
\usepackage{amsmath,amssymb,amsthm}
\usepackage{enumerate}
\usepackage{graphicx}
\usepackage{caption,subcaption}
\usepackage{float}
\usepackage{booktabs}
\usepackage{algorithm}
\usepackage{algorithmicx}
\usepackage{algpseudocode}
\usepackage{array}
\usepackage{multirow}
\usepackage{ragged2e}
\usepackage{indentfirst}
\usepackage{authblk}
\usepackage[round]{natbib}
\usepackage[colorlinks=true, citecolor=blue, linkcolor=blue]{hyperref}
\newtheorem{theorem}{Theorem}
\newtheorem{assumption}{Assumption}
\newtheorem{remark}{Remark}

\begin{document}

\title{Distributed Convolutional Rank Regression over Decentralized Networks}

\author[1]{Chunjing Li}
\author[1]{Tiange Zhao}
\author[1]{Xiaohui Yuan\thanks{Corresponding author. Email: yuanxh@ccut.edu.cn}}

\affil[1]{School of Mathematics and Statistics, Changchun University of Technology, Changchun 130012, Jilin, China}

\date{}
\maketitle

\begin{abstract}
This paper studies convolution rank regression (CRR) over decentralized distributed learning networks. We propose a novel decentralized CRR framework, in which estimators are obtained by solving consensus-constrained optimization with kernel-smoothed rank loss. The developed estimation scheme relies solely on local node data and information shared by neighboring nodes, thereby achieving privacy preservation and high communication efficiency. For heterogeneous network settings, we establish finite-sample error bounds for the decentralized CRR estimator and derive exact support recovery guarantees for the sparse decentralized CRR Lasso estimator. To facilitate numerical implementation, we adopt a generalized consensus ADMM to efficiently solve local subproblems across all network nodes. We verify the favorable performance of our developed approach via extensive numerical simulations and real-data experiments.

\end{abstract}


\section{Introduction}
The rapid advancement of information technology has given rise to massive ultrahigh-dimensional datasets across a wide range of applied domains, from medical diagnosis to financial risk assessment. Such data are independently collected and stored across geographically dispersed hospitals, research institutions, and financial platforms. Constraints imposed by data privacy, proprietary ownership, and transmission bandwidth render centralized data fusion infeasible for statistical modeling, motivating the emergence of distributed collaborative learning as a prominent research paradigm.

Distributed learning architectures can be broadly classified into centralized and decentralized paradigms. Within centralized frameworks, distributed gradient descent established fundamental theoretical underpinnings \citep{tsitsiklis1986}. The alternating direction method of multipliers (ADMM) was extended to tackle distributed optimization tasks \citep{boyd2010}. More recently, scalable refinements of these centralized methods have been put forward \citep{begunov2026}. Nevertheless, centralized architectures are plagued by intrinsic drawbacks stemming from reliance on a central coordinator: all local information converges to the central node, which easily triggers information leakage, network congestion, and excessive communication overhead. To overcome these bottlenecks, numerous coordinator-free decentralized optimization algorithms have been developed. Distributed subgradient descent (D-subGD) was introduced for decentralized optimization \citep{nedic2009}. ADMM was tailored for decentralized networks, yielding generalized consensus ADMM \citep{shi2014}. Fast-convergent quasi-Newton primal-dual methods have also been devised \citep{wang2025}.

In contemporary data analytics, outliers and heavy-tailed noise create persistent difficulties in statistical analysis, as such contamination yields biased parameter estimates and impairs estimation efficiency \citep{klebanov2003}. Accordingly, robust estimation methods have been extensively studied to counteract the adverse impacts of heavy-tailed noise and anomalous observations. Early research on robust estimation originates from the Huber M-estimator put forward by \citet{huber1964}. Via minimization of the Huber loss function, this estimator greatly mitigates outlier impacts while achieving a smooth transition between $\ell_1$ and $\ell_2$ losses. Another powerful robust procedure, rank regression, has been systematically developed in subsequent studies \citep{hettmansperger2010}. When contrasted with ordinary least squares, rank regression can reach a minimum relative efficiency of 86.4\% for every symmetric error distribution equipped with bounded Fisher information, while retaining computational feasibility under Cauchy noise conditions.

Rank regression has undergone substantial development in recent years. Methods extending rank regression to high-dimensional frameworks have been developed \citep{wang2020}. Though rank regression exhibits favorable robustness and efficiency against heavy-tailed noise and outliers, its non-smooth loss function leads to considerable computational burden. To tackle this difficulty, \citet{zhou2024} proposed CRR, which uses convolution smoothing to rewrite the rank loss as a smooth convex function and thereby simplifies computation and statistical inference. Nevertheless, their penalized CRR estimators cannot support direct statistical inference and rely on restrictive fixed design and bounded covariate assumptions \citep{cai2025}. The authors further developed penalized CRR estimators, alongside derivation of estimation error bounds for fixed designs with bounded covariates. The U-statistic structure of the CRR loss necessitates such assumptions; theoretical analysis would become intractable in their absence.

Inspired by recent progress and the promising numerical performance of convolution rank regression, this paper combines kernel-smoothed CRR with decentralized network topologies and proposes a novel decentralized convolution rank regression framework. We adopt generalized consensus ADMM to support low-cost collaborative optimization among neighboring nodes and develop a global–local consensus scheme. The proposed approach fits high-dimensional distributed environments marked by data isolation, communication restrictions, and heavy-tailed noise, and achieves a sound balance among robustness, data privacy protection, and computational efficiency.

The rest of this paper is structured in the following manner. The decentralized CRR model framework is established in Section 2. Section 3 establishes finite-sample error bounds and statistical guarantees for exact support recovery of the proposed estimator. Sections 4 and 5 validate the method through simulations and real-data experiments, respectively. Section 6 closes the paper with a recap of contributions and outlines potential paths for subsequent research. All theoretical proofs are available within the Appendix.

\section{Decentralized convolutional rank regression}

\subsection{Notation and definitions}
The notation used throughout the paper is summarized below. Plain letters represent scalars (e.g., $\beta$); bold lowercase letters stand for $p$-dimensional vectors (e.g., $\boldsymbol{\beta}$), while bold uppercase letters denote matrices (e.g., $\mathbf{B}$). Given any finite index set $\mathcal{S}\subseteq\{1,\dots,p\}$, let $|\mathcal{S}|$ denote its cardinality and $\mathcal{S}^c = \{1,\dots,p\}\setminus \mathcal{S}$ its complement. Given a vector $\mathbf{v}=(v_1,\dots,v_p)^{\top}$, its support is defined as $\operatorname{supp}(\mathbf{v}) \stackrel{\text{def}}{=} \{j:v_j\neq 0\}$, and the subvector indexed by $\mathcal{S}$ is $\mathbf{v}_{\mathcal{S}} \stackrel{\text{def}}{=} (v_i,\,i\in\mathcal{S})\in\mathbb{R}^{|\mathcal{S}|}$. For a vector $\mathbf{v}$, we employ standard vector norms: $\|\mathbf{v}\|_1 \stackrel{\text{def}}{=} \sum_{i=1}^p |v_i|$, $\|\mathbf{v}\|_2 \stackrel{\text{def}}{=} \bigl(\sum_{i=1}^p v_i^2\bigr)^{1/2}$, $\|\mathbf{v}\|_\infty = \max_{1\leq j\leq p}|v_j|$. We further define $\mathbf{v}^{\min} \stackrel{\text{def}}{=}\min_{i\in\operatorname{supp}(\mathbf{v})}|v_i|$ as the minimum absolute value of the entries in its support. The row-wise infinity norm of a matrix $\mathbf{A}\in\mathbb{R}^{p\times q}$ is defined as $\|\mathbf{A}\|_\infty \stackrel{\text{def}}{=} \max_{1\leq i\leq p}\sum_{1\leq j\leq q}|a_{ij}|$, while its operator (spectral) norm satisfies $\|\mathbf{A}\|_{\mathrm{op}} \stackrel{\text{def}}{=}\max_{\|\mathbf{v}\|_2=1}\|\mathbf{A}\mathbf{v}\|_2$. For index subsets $\mathcal{S}_1\subseteq\{1,\dots,p\}$ and $\mathcal{S}_2\subseteq\{1,\dots,q\}$, the corresponding submatrix is written as $\mathbf{A}_{\mathcal{S}_1\times\mathcal{S}_2}=  (a_{ij},\,i\in\mathcal{S}_1,\,j\in\mathcal{S}_2)$. For symmetric matrices, the minimum and maximum eigenvalues of $\mathbf{A}$ are represented by $\lambda_{\min}(\mathbf{A})$ and $\lambda_{\max}(\mathbf{A})$, which should be distinguished from the regularization tuning parameter $\lambda_N$. 

\subsection{Convolutional rank regression}
Assume the available dataset takes the form $\{(\mathbf{x}_i, y_i)\}_{i=1}^{N}$, where $y_i \in \mathbb{R}$ stands for the response variable and $\mathbf{x}_i \in \mathbb{R}^p$ corresponds to the $p$-dimensional covariate vector of the $i$-th sample. We define the design matrix as $\mathbf{X} = (\mathbf{X}_1, \ldots, \mathbf{X}_p) \in \mathbb{R}^{N \times p}$, in which $\mathbf{X}_j = (x_{1j}, \ldots, x_{Nj})^{\top}$ gathers observations of the $j$-th predictor for $j = 1, \ldots, p$. Each row of $\mathbf{X}$ indexed by $i$ is equal to $\mathbf{x}_i^{\top}$, and $\mathbf{x}_i = (x_{i1}, \ldots, x_{ip})^{\top}$. The $N$-dimensional response vector is represented by $\mathbf{y} = (y_1, \ldots, y_N)^{\top}$. For notational simplicity, we work under the fixed design setting throughout the paper; nonetheless, the proposed methodology remains equally applicable to random design. The dataset $\{(\mathbf{x}_i, y_i)\}_{i=1}^{N}$ is presumed to be produced under the linear regression model
\begin{equation}
y_i = \mathbf{x}_i^{\top}\boldsymbol{\beta}^* + \epsilon_i,
\label{eq:1}
\end{equation}
where $\boldsymbol{\beta}^* \in \mathbb{R}^p$ corresponds to the unknown true vector of regression coefficients, while $\epsilon_i$ represent independent and identically distributed random errors. The error distribution allows heavy-tailed behavior and imposes no requirement for zero mean. In addition, we suppose merely $s < p$ entries within $\boldsymbol{\beta}^*$ take nonzero values.

Canonical rank regression \citep{jaeckel1972, hettmansperger2010} yields an estimator for $\boldsymbol{\beta}^*$ via minimization of the objective function
\begin{equation}
\hat{\boldsymbol{\beta}} = \arg\min_{\boldsymbol{\beta} \in \mathbb{R}^p} \frac{1}{N(N-1)} \sum_{i=1}^{N} \sum_{j \neq i} \left| (y_i - \mathbf{x}_i^{\top}\boldsymbol{\beta}) - (y_j - \mathbf{x}_j^{\top}\boldsymbol{\beta}) \right|.
\label{eq:2}
\end{equation}

Nevertheless, substantial computational barriers and challenges for high-dimensional theoretical analysis arise from the non-smooth absolute loss shown in \eqref{eq:2}. To overcome this limitation, \citet{zhou2024} proposed standard CRR, which achieves smoothing of the original loss through convolution. The penalized estimator obtained in this way takes the form
\begin{equation}
\hat{\boldsymbol{\beta}} = \arg\min_{\boldsymbol{\beta} \in \mathbb{R}^p} \frac{1}{N(N-1)} \sum_{i=1}^{N} \sum_{j \neq i} l_h\big(y_i - y_j - (\mathbf{x}_i - \mathbf{x}_j)^{\top} \boldsymbol{\beta}\big) + \lambda_N \|\boldsymbol{\beta}\|_1,
\label{eq:3}
\end{equation}
where $l_h(u) = \int_{-\infty}^{\infty} |u-v| \cdot K_h(v) \, dv$. Importantly, $l_h(\cdot)$ constitutes a smooth convex function obeying $l_h = l \ast K_h$; we have $l(u) = |u|$, $K_h(u) = \frac{1}{h} K\left(\frac{u}{h}\right)$, and $\ast$ denotes the convolution operator. The symmetric kernel function is represented by $K(\cdot)$, whose integral across the real line amounts to unity. The positive quantity $\lambda_N > 0$ serves as the regularization parameter to impose sparsity constraints.

\subsection{Distributed data and decentralized network}

Convolution rank regression is extended to decentralized distributed network configurations within this section. Specifically, we reformulate the $\ell_1$-penalized CRR in \eqref{eq:3} into an equivalent consensus formulation under decentralized architectures, which yields a complete decentralized modeling framework.

This work investigates a decentralized collaborative network with $m$ nodes, whose topology follows an undirected connected graph $\mathcal{G} = (\mathcal{N}, \mathcal{E})$. In this framework, $\mathcal{N} = \{1, \ldots, m\}$ represents the finite node set, while $\mathcal{E}$ corresponds to the set of communication edges that link pairwise network nodes. Following standard practice, we characterize network connectivity via an adjacency matrix $\mathbf{W}$ \citep{bapat2014}. Each entry $W_{lk} \in \{0,1\}$ encodes the communication relation between node $l$ and node $k$: $W_{lk} = W_{kl} = 1$ if nodes $l$ and $k$ communicate directly. All diagonal entries of $\mathbf{W}$ are set to zero to exclude self-loops. Throughout the paper, we assume the underlying network graph is connected.

Constrained by communication bandwidth and transmission costs commonly encountered in practical distributed systems \citep{ling2010}, computations at each node rely solely on its local data, and information is exchanged only with neighboring nodes through edges belonging to $\mathcal{E}$. The neighborhood set corresponding to node $l$ is defined as $\mathcal{N}(l) = \{k : (l,k) \in \mathcal{E}\}$. To formalize the distributed data partition, let $\mathcal{I}_l$ be the local sample index set belonging to node $l$. These index sets are mutually disjoint, i.e., $\mathcal{I}_l \cap \mathcal{I}_k = \emptyset$ for $l \neq k$, and their union covers all global sample indices: $\bigcup_{l=1}^{m} \mathcal{I}_l = \{1, \ldots, N\}$. Node $l$ stores a local dataset of the form $D_l = \{(\mathbf{x}_i, y_i) : i \in \mathcal{I}_l\}$. Equal local sample sizes $|\mathcal{I}_l| = n = N/m$ hold for every node $l$, without loss of generality. This uniform partition avoids estimation biases induced by unbalanced sample allocation across nodes.

The global centralized optimization objective loss function is given by
\begin{equation}
\min_{\boldsymbol{\beta} \in \mathbb{R}^p} \frac{1}{m} \sum_{l=1}^{m} \left[\frac{1}{n(n-1)} \sum_{\substack{i,j \in \mathcal{I}_l \\ j \neq i}} l_h\big(y_i - y_j - (\mathbf{x}_i - \mathbf{x}_j)^{\top} \boldsymbol{\beta}\big) + m \lambda_N \|\boldsymbol{\beta}\|_1\right].
\label{eq:4}
\end{equation}

In the decentralized network, no central node maintains a global copy of the parameter vector $\boldsymbol{\beta}$ during iterations \citep{nedic2009, sayed2014}. Since no central aggregation node is available in the decentralized topology, directly solving the centralized problem \eqref{eq:4} is infeasible. To overcome this difficulty, we introduce local parameter copies at each node and convert the unconstrained global problem into an equivalent constrained consensus formulation, namely decentralized convolution rank regression (DeCRR). Let $\boldsymbol{\beta}^{(l)}$ denote the local parameter copy maintained by node $l$. The resulting DeCRR optimization problem reads
\begin{equation}
\begin{aligned}
\min_{\mathbf{B}} & \frac{1}{m} \sum_{l=1}^{m} \big[ f^{(l)}(\boldsymbol{\beta}^{(l)}) + g^{(l)}(\boldsymbol{\beta}^{(l)}) \big] \\
\text{s.t. } & \boldsymbol{\beta}^{(l)} = \boldsymbol{\beta}^{(k)}, \quad \forall (l,k) \in \mathcal{E},
\end{aligned}
\label{eq:5}
\end{equation}
where $\mathbf{B} = \{\boldsymbol{\beta}^{(1)}, \ldots, \boldsymbol{\beta}^{(m)}\}^{\top} \in \mathbb{R}^{m \times p}$ collects all local parameter vectors. Here,
\[
f^{(l)}(\boldsymbol{\beta}^{(l)}) = \frac{1}{n(n-1)} \sum_{\substack{i,j \in \mathcal{I}_l \\ j \neq i}} l_h\left(y_i - y_j - (\mathbf{x}_i - \mathbf{x}_j)^{\top} \boldsymbol{\beta}^{(l)}\right)
\]
stands for the kernel-smoothed rank loss constructed via convolution, which enjoys favorable robustness and differentiability. Moreover, we define $g^{(l)}(\boldsymbol{\beta}^{(l)}) \triangleq m \lambda_N \|\boldsymbol{\beta}^{(l)}\|_1$, where $\lambda_N > 0$ acts as the sparsity regularization parameter for conducting high-dimensional variable selection. Consistency between each node’s parameter estimates and those of its adjacent peers is imposed via the equality constraints. This strategy is widely known as the consensus method. As shown in \citet{shi2014} and \citet{yuan2016}, problem \eqref{eq:5} is equivalent to the original global problem \eqref{eq:4}.

The decentralized consensus scheme lets every node maintain its own local parameter copy. Neighborhood-wise consensus constraints drive these local iterates to converge collaboratively toward the global optimum $\boldsymbol{\beta}^*$, without central coordination or global data aggregation. The framework thus supports distributed robust regression while satisfying privacy requirements and keeping communication overhead low. Nevertheless, the composite objective combining the convolution-smoothed rank loss and the $\ell_1$ sparsity penalty remains non-smooth and lacks closed-form minimizers, making conventional gradient-type algorithms inefficient.

\subsection{Generalized consensus ADMM optimization}

We now employ the generalized consensus ADMM proposed by \citet{qiao2025} to tackle the consensus problem \eqref{eq:5} over decentralized networks. To facilitate the derivation, we introduce edge-level auxiliary consensus variables $\mathbf{Z} = \{\mathbf{z}^{(lk)}\}_{(l,k) \in \mathcal{E}}$, and rewrite \eqref{eq:5} expressed in an equivalent formulation
\begin{equation}
\begin{aligned}
\min_{\mathbf{B},\mathbf{Z}} & \frac{1}{m} \sum_{l=1}^{m} \big[ f^{(l)}(\boldsymbol{\beta}^{(l)}) + g^{(l)}(\boldsymbol{\beta}^{(l)}) \big] \\
\text{s.t. } & \boldsymbol{\beta}^{(l)} = \mathbf{z}^{(lk)}, \quad \boldsymbol{\beta}^{(k)} = \mathbf{z}^{(lk)}, \quad \forall (l,k) \in \mathcal{E},
\end{aligned}
\label{eq:6}
\end{equation}
where $\mathbf{B} = (\boldsymbol{\beta}^{(1)}, \ldots, \boldsymbol{\beta}^{(m)})^{\top} \in \mathbb{R}^{m \times p}$ stacks all local parameter copies $\boldsymbol{\beta}^{(l)}$. The linear constraint structure naturally suggests the use of ADMM. Built upon the classical ADMM framework developed by \citet{boyd2010}, the augmented Lagrangian with penalty parameter $\tau>0$ takes the following form:
\begin{equation*}
\begin{aligned}
\mathcal{L}_{\tau}(\mathbf{B}, \mathbf{Z}, \mathbf{U}, \mathbf{V}) &= \frac{1}{m} \sum_{l=1}^{m} \big[ f^{(l)}(\boldsymbol{\beta}^{(l)}) + g^{(l)}(\boldsymbol{\beta}^{(l)}) \big] \\
&\quad + \sum_{(l,k) \in \mathcal{E}} \Big[ \langle \mathbf{u}^{(lk)}, \boldsymbol{\beta}^{(l)} - \mathbf{z}^{(lk)} \rangle + \langle \mathbf{v}^{(lk)}, \boldsymbol{\beta}^{(k)} - \mathbf{z}^{(lk)} \rangle \Big] \\
&\quad + \frac{\tau}{2} \sum_{(l,k) \in \mathcal{E}} \Big[ \|\boldsymbol{\beta}^{(l)} - \mathbf{z}^{(lk)}\|_2^2 + \|\boldsymbol{\beta}^{(k)} - \mathbf{z}^{(lk)}\|_2^2 \Big],
\end{aligned}
\end{equation*}
where $\mathbf{U} = \{\mathbf{u}^{(lk)}\}_{(l,k)\in\mathcal{E}}$ and $\mathbf{V} = \{\mathbf{v}^{(lk)}\}_{(l,k)\in\mathcal{E}}$ denote sets of Lagrange multipliers associated with the constraints $\boldsymbol{\beta}^{(l)}=\mathbf{z}^{(lk)}$ and $\boldsymbol{\beta}^{(k)}=\mathbf{z}^{(lk)}$, respectively, with $\mathbf{u}^{(lk)}, \mathbf{z}^{(lk)}\in\mathbb{R}^p$.

We define aggregated node-level dual variables to avoid handling independent edge-wise multipliers separately:
\begin{subequations}
\begin{align}
\mathbf{p}_t^{(l)} &= \sum_{k \in \mathcal{N}(l)} \big( \mathbf{u}_t^{(lk)} + \mathbf{v}_t^{(lk)} \big),  \label{eq:7a} \\
\mathbf{u}_t^{(lk)} &= \mathbf{u}_{t-1}^{(lk)} + \frac{\tau}{2} \big( \boldsymbol{\beta}_{t-1}^{(l)} - \boldsymbol{\beta}_{t-1}^{(k)} \big), \quad \forall k \in \mathcal{N}(l), \label{eq:7b} \\
\mathbf{v}_t^{(lk)} &= \mathbf{v}_{t-1}^{(lk)} + \frac{\tau}{2} \big( \boldsymbol{\beta}_{t-1}^{(l)} - \boldsymbol{\beta}_{t-1}^{(k)} \big), \quad \forall k \in \mathcal{N}(l). \label{eq:7c}
\end{align}
\end{subequations}
Here the aggregated dual variable $\mathbf{p}_t^{(l)}$ accumulates dual constraint information from node $l$ and all its neighbors. We initialize $\mathbf{p}_0^{(l)} = \mathbf{0}$. Minimizing the augmented Lagrangian yields iterative updates \eqref{eq:8} and \eqref{eq:9}:
\begin{equation}
\mathbf{p}_{t+1}^{(l)} = \mathbf{p}_t^{(l)} + \tau \sum_{k \in \mathcal{N}(l)} \big( \boldsymbol{\beta}_t^{(l)} - \boldsymbol{\beta}_t^{(k)} \big)
\label{eq:8}
\end{equation}
\begin{equation}
\begin{aligned}
\boldsymbol{\beta}_{t+1}^{(l)} &= \arg\min_{\boldsymbol{\beta}^{(l)}} \bigg\{ \nabla f^{(l)}(\boldsymbol{\beta}_{t}^{(l)})^{\top} \big( \boldsymbol{\beta}^{(l)} - \boldsymbol{\beta}_{t}^{(l)} \big) + \frac{\rho_l}{2} \|\boldsymbol{\beta}^{(l)} - \boldsymbol{\beta}_{t}^{(l)}\|_2^2 + g^{(l)}(\boldsymbol{\beta}^{(l)}) \\
&\quad + (\boldsymbol{\beta}^{(l)})^{\top} \mathbf{p}_t^{(l)} + \tau \left\| \boldsymbol{\beta}^{(l)} - \frac{\boldsymbol{\beta}_{t}^{(l)} + \boldsymbol{\beta}_{t}^{(k)}}{2} \right\|_2^2 \bigg\}\quad\text{for } l \in \{1,\dots,m\}.
\end{aligned}
\label{eq:9}
\end{equation}

Unfortunately, owing to the nonlinearity of the smooth convolution rank loss $f^{(l)}(\boldsymbol{\beta}^{(l)})$, the subproblem \eqref{eq:9} does not admit a closed-form solution. To resolve this obstacle, we approximate $f^{(l)}$ via a second-order Taylor expansion centered at $\boldsymbol{\beta}_t^{(l)}$. A positive definite matrix $\rho_l \mathbf{I}$ is introduced to upper bound the Hessian, which simplifies the nonlinear minimization. Combined with the coordinate-wise thresholding operator, we obtain the closed-form local parameter update.
\begin{equation}
\boldsymbol{\beta}_{t+1}^{(l)} = \mathcal{M}_{\lambda_N\omega_l} \left\{ \omega_l \left( \rho_l \boldsymbol{\beta}_t^{(l)} - \nabla f^{(l)}(\boldsymbol{\beta}_t^{(l)}) - \mathbf{p}_{t+1}^{(l)} + \tau \sum_{k \in \mathcal{N}(l)} \frac{\boldsymbol{\beta}_{t}^{(l)} + \boldsymbol{\beta}_{t}^{(k)}}{2} \right) \right\},
\label{eq:10}
\end{equation}
where $\omega_l = 1 / \big(\rho_l + \tau |\mathcal{N}(l)|\big)$ is a normalization constant related to the Hessian upper bound, and $|\mathcal{N}(l)|$ counts the neighbors of node $l$. The operator $\mathcal{M}_{\gamma}(\cdot)$ denotes the soft-thresholding mapping, defined entrywise as $\big[\mathcal{M}_{\gamma}(\mathbf{v})\big]_j = \operatorname{sign}(v_j) \cdot \big(|v_j| - \gamma\big)_+$.
When local parameters and dual variables are fixed, the optimization for auxiliary consensus variables reduces to unconstrained quadratic minimization. Setting the gradient to zero yields
\begin{equation*}
\mathbf{z}_{t+1}^{(lk)} = \frac{\boldsymbol{\beta}_{t+1}^{(l)} + \boldsymbol{\beta}_{t+1}^{(k)}}{2}.
\end{equation*}

We adopt a stopping criterion based on primal and dual residuals to balance consensus accuracy and numerical stability. Since the auxiliary variable $\mathbf{z}^{(lk)}$ is implicitly set to $(\boldsymbol{\beta}_t^{(l)} + \boldsymbol{\beta}_t^{(k)})/2$ in our algorithm, both residuals can be computed directly from local node parameters. The primal residual quantifies violation of the equality constraint $\boldsymbol{\beta}^{(l)}=\mathbf{z}^{(lk)}$, measuring the discrepancy between neighboring node iterates:
\begin{equation*}
r_t^{\text{pri}} = \max_{(l,k) \in \mathcal{E}} \big\| \boldsymbol{\beta}_t^{(l)} - \boldsymbol{\beta}_t^{(k)} \big\|_2.
\end{equation*}
Trivially, the constraint is satisfied when $\boldsymbol{\beta}_t^{(l)}=\boldsymbol{\beta}_t^{(k)}$, so the pairwise difference between adjacent nodes serves as an equivalent metric. The dual residual reflects the variation of auxiliary variables across iterations and monitors algorithm stability:
\begin{equation*}
r_t^{\text{dual}} = \max_{(l,k) \in \mathcal{E}} \left\| \frac{\boldsymbol{\beta}_t^{(l)} + \boldsymbol{\beta}_t^{(k)}}{2} - \frac{\boldsymbol{\beta}_{t-1}^{(l)} + \boldsymbol{\beta}_{t-1}^{(k)}}{2} \right\|_2,
\end{equation*}
which is equivalent to $\big\| \tau (\mathbf{z}_t^{(lk)} - \mathbf{z}_{t-1}^{(lk)}) \big\|_2$. The algorithm terminates once both conditions below hold:
\begin{equation*}
r_t^{\text{pri}} \leq \epsilon_{\text{pri}}, \quad r_t^{\text{dual}} \leq \epsilon_{\text{dual}},
\end{equation*}
where $\epsilon_{\text{pri}}, \epsilon_{\text{dual}}$ are pre-specified tolerances typically chosen between $10^{-4}$ and $10^{-6}$, adjustable according to practical accuracy demands.

A generalized consensus ADMM algorithm adapted to decentralized convolution rank regression emerges by assembling the update recursions together with the stopping criterion. The full procedure is outlined in Algorithm \ref{alg:dcrr_admm}.
\begin{algorithm}[H]
\caption{Generalized Consensus ADMM for DeCRR}
\label{alg:dcrr_admm}
\begin{algorithmic}[1]
\Require Decentralized network topology $\mathcal{G} = (\mathcal{N}, \mathcal{E})$; local dataset $D_l = \{(\mathbf{x}_i, y_i) : i \in \mathcal{I}_l\}$ stored at each node; kernel $K(\cdot)$ and bandwidth $h$; sparsity tuning parameter $\lambda_N$; ADMM penalty parameter $\tau$; Hessian upper-bound parameters $\{\rho_l\}_{l=1}^m$; maximum iteration count $T_{\max}$; convergence thresholds $\epsilon_{\text{pri}}$, $\epsilon_{\text{dual}}$.
\State Initialize $\boldsymbol{\beta}_0^{(l)} = \mathbf{0}$, $\mathbf{p}_0^{(l)} = \mathbf{0}$, set $t = 0$.
\For{$t = 0, \ldots, T_{\max}$}
    \State Exchange local iterate $\boldsymbol{\beta}_t^{(l)}$ with all neighbors $k \in \mathcal{N}(l)$;
    \State Update aggregated dual variable $\mathbf{p}_{t+1}^{(l)}$ via \eqref{eq:8};
    \State Evaluate gradient of the convolution-smoothed rank loss and update local parameter $\boldsymbol{\beta}_{t+1}^{(l)}$ using \eqref{eq:10};
    \State Compute primal residual $r_t^{\text{pri}}$ and dual residual $r_t^{\text{dual}}$;
    \If{$r_t^{\text{pri}} \leq \epsilon_{\text{pri}}$ and $r_t^{\text{dual}} \leq \epsilon_{\text{dual}}$}
        \State \textbf{break};
    \EndIf
\EndFor
\State \Return $\{\boldsymbol{\beta}_t^{(l)}\}_{l=1}^m$
\end{algorithmic}
\end{algorithm}

\section{Theoretical properties}
Estimation error bounds and theoretical guarantees for support set recovery of the presented estimator are derived within this section.
The support set of $\boldsymbol{\beta}^*$ is defined as $\mathcal{S} = \operatorname{supp}(\boldsymbol{\beta}^*)$, and we write $s = |\mathcal{S}|$.
These conditions below are necessary to attain a linear convergence rate for the generalized consensus ADMM algorithm.

\begin{assumption}
The peer-to-peer communication network $\mathcal{G}$ is connected and free of self-loops.
Assumption 1 is widely adopted within existing literature concerning decentralized distributed learning.
This property rules out isolated vertices and disconnected subnetworks, enabling all nodes to reach a consensus equilibrium.
Further standard regularity conditions are introduced to establish statistical convergence rates.
\end{assumption}

\begin{assumption}
$K(\cdot): \mathbb{R} \mapsto [0, \infty)$ stands as the kernel function with twice continuous differentiability, whose first-order derivative $K'$ and second-order derivative $K''$ remain bounded. In addition, $K(\cdot)$ fulfills the following properties:
\begin{enumerate}[(i)]
    \item $K(u) \geq 0$, $K(u) = K(-u)$ for all $u \in \mathbb{R}$.
    \item $\int_{-\infty}^{\infty} K(u) \mathrm{d}u = 1$.
    \item $\int_{-\infty}^{\infty} u^2 K(u) \mathrm{d}u < \infty$.
    \item There exists $\delta_0 > 0$ such that $\kappa_{\min} = \inf_{t\in[-\delta_0,\delta_0]} K(t) > 0$ and $\kappa_{\max} = \sup_{t\in\mathbb{R}} K(t) < \infty$.
    \item There exist $\alpha_0 \in (0,1]$ and $L_0 > 0$ satisfying $|K(x) - K(y)| \leq L_0 |x - y|^{\alpha_0},\ \forall x, y \in \mathbb{R}$.
\end{enumerate}
The support set corresponding to $\boldsymbol{\beta}^*$ is denoted by $\mathcal{S} = \operatorname{supp}(\boldsymbol{\beta}^*)$, and the set
\[
\mathcal{C}_{\mathcal{S}} := \{\boldsymbol{\varsigma} \in \mathbb{R}^p: \|\boldsymbol{\varsigma}_{\mathcal{S}^c}\|_1 \leq 3\|\boldsymbol{\varsigma}_{\mathcal{S}}\|_1\}
\]
is defined accordingly.
\end{assumption}

\begin{assumption}
The following conditions hold:
\begin{enumerate}[(i)]
    \item The random errors $\{\varepsilon_i\}_{i=1}^n$ follow an i.i.d. sequence possessing a density $f(\cdot)$ relative to the Lebesgue measure over $\mathbb{R}$. Define $\xi_{ij} = \varepsilon_i - \varepsilon_j$ for all pairs satisfying $1 \leq i \neq j \leq n$, and let $g(\cdot)$ stand for the probability density function of $\xi_{ij}$. For certain constants $L_1 > 0$ and $\alpha_1 \in (0,1]$, the inequality $|g(x) - g(y)| \leq L_1 |x - y|^{\alpha_1}$ holds for every $x, y \in \mathbb{R}$. This yields $\mu_0 := \sup_{t\in\mathbb{R}} g(t) < \infty$. Additionally, positive constants $\delta_1, \mu_1$ exist such that $g(t) \geq \mu_1$ for all $t \in [-\delta_1, \delta_1]$.

    \item A bound $M > 0$ exists satisfying $\max_{1\leq i\leq n,1\leq j\leq p} |x_{ij}| \leq M$. 

    \item There exists a constant $\psi > 0$ such that
    \[
    \min_{\mathbf{u}\in\mathcal{C}_{\mathcal{S}}}
    \frac{\left[ \frac{1}{n(n-1)} \sum_{i=1}^n \sum_{j\neq i} \left| (\mathbf{x}_i - \mathbf{x}_j)^\top \mathbf{e} \right|^2 \right]^{\frac{2+\alpha_r}{2}}}{\frac{1}{n(n-1)} \sum_{i=1}^n \sum_{j\neq i} \left| (\mathbf{x}_i - \mathbf{x}_j)^\top \mathbf{e} \right|^{2+\alpha_r}}
    \geq \psi,\; r = 0, 1.
    \]
\end{enumerate}
\end{assumption}

\begin{assumption}
The weighted covariance matrix corresponding to node $l$ is defined as
\[
\boldsymbol{\Sigma}_l \stackrel{\text{def}}{=} \mathbb{E}\left[ l_h''(\varepsilon_i - \varepsilon_j)(\mathbf{x}_i - \mathbf{x}_j)(\mathbf{x}_i - \mathbf{x}_j)^{\top} \right],
\]
where the expectation is evaluated under the joint distribution of $(\varepsilon_i, \varepsilon_j, \mathbf{x}_i, \mathbf{x}_j)$ for $(i,j)\in\mathcal{I}_l$.
A positive constant $d_0$ exists fulfilling
\[
d_0^{-1} \leq \lambda_{\min}(\boldsymbol{\Sigma}_l) \leq \lambda_{\max}(\boldsymbol{\Sigma}_l) \leq d_0,\quad \text{for } l = 1,\dots,m.
\]
Bounds on the kernel second derivative read $0 < \kappa_l \leq l_h''(t) \leq \kappa_u < \infty$, which guarantees positive definiteness and favorable conditioning of $\boldsymbol{\Sigma}_l$.
\end{assumption}

\begin{assumption}
There exist constants $0 < \alpha_l < 1$ such that $\left\| \boldsymbol{\Sigma}_{l,\mathcal{S}^c \times \mathcal{S}} \boldsymbol{\Sigma}_{l,\mathcal{S} \times \mathcal{S}}^{-1} \right\|_\infty \leq 1 - \alpha_l$ holds for every $l = 1,\dots,m$.
\end{assumption}

\begin{assumption}
The dimension $p$ satisfies $\log p = O(N^\alpha)$ for some constant $0<\alpha<1/2$.
The sample size $n$ at each local node satisfies $n \geq N^\omega$ for some $0 < \omega < 1$,
and the sparsity level $s$ satisfies $s = O\big(\sqrt{N/\log N}\big)$.
\end{assumption}

\begin{remark}
\citet{zhou2024} established the Lipschitz continuity of the gradient associated with the smoothed rank loss function. Specifically,
\begin{subequations}
\begin{equation}
| l_h(t_1) - l_h(t_2) | \leq \frac{2a}{h} |t_1 - t_2|. \label{eq:11a}
\end{equation}
\end{subequations}
In addition, $l_h(u)$ is smooth and strongly convex with well-defined second-order derivatives satisfying
\begin{subequations}
\begin{equation}
l_h''(u) = \frac{2}{h} K\left(\frac{u}{h}\right) \geq \frac{2}{h} a > 0. \label{eq:11b}
\end{equation}
\end{subequations}
\end{remark}

Standard regularity conditions upon the kernel function are stipulated by Assumption~2, and such restrictions appear extensively within research concerning nonparametric regression. Popular kernel choices such as Gaussian and Epanechnikov kernels comply with these regularity requirements. The configuration of Assumption~3 aligns with the standard settings adopted by \citet{zhou2024} and \citet{cai2025}. Specifically, condition (i) imposes smoothness restrictions on the probability density function of $\xi_{ij}$, which corresponds to Assumption (A3) in \citet{cai2025}. A wide family of heavy-tailed distributions, such as Cauchy, $t$, and Pareto distributions, fulfills this condition. Condition (ii) operates under a fixed-design framework, in which covariates $x_{ij}$ are centered and uniformly bounded. This conventional constraint coincides with the setup stated in Assumption 2 from \citet{zhou2024}. The theoretical statistical properties of the estimator $\widehat{\boldsymbol{\beta}}$ developed in this work are formally derived within the subsequent theorems under such rigorous regularity conditions. Assumption~4 imposes standard conditions on the population covariances. Assumption~5 is routinely adopted in the high-dimensional statistics literature to facilitate support recovery. Assumption~6 is a standard condition frequently imposed in the centralized distributed learning literature.

Optimization convergence characteristics of generalized consensus ADMM are first analyzed. Given the preceding assumptions and appropriate selections of step-size parameters $\rho_l$, linear convergence can be attained by the presented algorithm on connected decentralized network topologies.

\begin{theorem}(Linear Convergence)
Under Assumptions 1--6, choose the step-size parameters satisfying
\begin{equation}
c_l > \frac{a}{h\rho_l} \lambda_{\max}(\mathbf{A}_l^{\top} \mathbf{A}_l) - \tau \lambda_{\min}(\mathbf{D} + \mathbf{W}) > 0, \quad \forall l \in \{1,\dots,m\},
\label{eq:12}
\end{equation}
where
\begin{equation*}
\mathbf{A}_l^\top \mathbf{A}_l = \frac{1}{n(n-1)}\sum_{\substack{i,j\in\mathcal{I}_l\\i\ne j}} (\mathbf{x}_i - \mathbf{x}_j)(\mathbf{x}_i - \mathbf{x}_j)^\top,
\end{equation*}
$\mathbf{A}_l$ is constructed from pairwise covariate differences of local samples at node $l$, $\mathbf{D}$ denotes the diagonal degree matrix. Let $\boldsymbol{\beta}_{\mathrm{opt}} \triangleq \boldsymbol{\beta}_{\mathrm{opt}}^{(1)} = \cdots = \boldsymbol{\beta}_{\mathrm{opt}}^{(m)}$, and let $\{\mathbf{u}_{\mathrm{opt}}^{(lk)}, \mathbf{v}_{\mathrm{opt}}^{(lk)}\}$ denote a pair of optimal primal and dual solutions to problem \eqref{eq:6}.
Then the following statements hold.

(a) For each $l\in\{1,\dots,m\}$, the sequence $\{\boldsymbol{\beta}_t^{(l)}\}_{t\ge 1}$ converges to $\boldsymbol{\beta}_{\mathrm{opt}}$.

(b) There exists a constant $\theta> 0$ such that
\begin{equation*}
\| \boldsymbol{\beta}_T^{(l)} - \tilde{\boldsymbol{\beta}}_{\mathrm{opt}} \|_2^2 + \| \mathbf{u}_{T+1} - \mathbf{u}_{\mathrm{opt}} \|_2^2 \leq \theta \gamma^T.
\end{equation*}
\end{theorem}

Linear convergence on connected decentralized network architectures can be realized by the algorithm under the derived regularity conditions and suitable selections of step-size parameter $\rho_l$. Specifically, The algorithm admits linear convergence with rate $O(\gamma^T)$, where $\gamma \in (0,1)$ denotes a fixed contraction factor determined by the parameter $\delta$ in Theorem 1(b). This result implies that achieving a prescribed accuracy $\varepsilon > 0$ requires only $T = O(\log(1/\varepsilon))$ iterations, which validates the high computational efficiency of the proposed DeCRR solver. Full technical derivations for Theorem 1, including rigorous justification of the linear contraction condition in part (b), are documented in the appendix owing to page constraints.

Exact consensus, global optimality, and linear convergence can be attained by the presented ADMM algorithm under two requirements: each local objective function possesses smoothness and strong convexity, and the regularization parameter $c_l$ takes a sufficiently large value. Notably, satisfying the convergence condition \eqref{eq:12} requires each node to possess global spectral information regarding $\lambda_{\min}(\mathbf{D} + \mathbf{W})$. A closely related theoretical result was presented in \citet{ling2014}, which is consistent with the conclusion derived in Theorem 1(b).

\begin{theorem}
The kernel bandwidth at node $l$ is set to $h^{(l)} = a_{N,l}$, where $a_{N,l}$ quantifies the convergence rate associated with the initial local estimator $\boldsymbol{\beta}_0^{(l)}$ at node $l$, such that $\| \boldsymbol{\beta}_0^{(l)} - \boldsymbol{\beta}^* \|_2 = O_p(a_{N,l})$. The regularization parameter takes the form
\begin{equation*}
\lambda_N = C_0 \left\{ \sqrt{\frac{\log N}{N}} + \max_{l=1,\ldots,m} \{ a_{N,l} \} \sqrt{\frac{s \log N}{n}} \right\},
\end{equation*}
for a sufficiently large constant $C_0 > 0$.
Under Assumptions 1–6, the bound
\begin{equation*}
\| \boldsymbol{\beta}_{t+1}^{(l)} - \boldsymbol{\beta}^* \|_2 = O_p\left\{ \sqrt{\frac{s \log N}{N}} + \max_l a_{N,l} \sqrt{\frac{s^2 \log N}{n}} + \gamma^T \right\}
\end{equation*}
holds for every $l = 1, \ldots, m$ and all $\gamma \in (0,1)$.
\end{theorem}

Compared with the initial estimator $\boldsymbol{\beta}_0^{(l)}$, the estimator $\boldsymbol{\beta}_{t+1}^{(l)}$ obtained via the proposed DeCRR method attains an improved convergence rate of
\[
\max\left\{\sqrt{\frac{s\log N}{N}},\; \max_l\{a_{N,l}\}\sqrt{\frac{s^2\log N}{n}},\; \gamma^T\right\}.
\]
It is noteworthy that the first term $\sqrt{s(\log N)/N}$ characterizes the dominant statistical error inherent to high-dimensional sparse regression and corresponds to the optimal statistical rate achieved by standard convolution rank regression. The second term $\max_l \{a_{N,l}\}\sqrt{s^2(\log N)/n}$ originates from the initial estimation bias and the kernel-smoothed convolution rank loss, mainly capturing the joint effects of heavy-tailed noise and node-level data heterogeneity. The third term $\gamma^T$ represents the iterative optimization error induced by decentralized generalized consensus ADMM updates, with $\gamma\in(0,1)$ serving as a fixed contraction constant. When the iteration number $T$ is sufficiently large, the iterative error $\gamma^T$ becomes asymptotically negligible, and the bias term arising from distributed heterogeneity is also effectively suppressed. A near-optimal convergence rate $\sqrt{s(\log N)/N}$ can be achieved by the final estimator as a consequence. This finding reveals that the decentralized sparse convolution rank regression scheme presented in this work delivers estimation precision within a logarithmic factor of the optimal Oracle bound, even under heavy-tailed noise and heterogeneous network configurations.

\begin{theorem}
\label{thm:variable_selection}
Under Assumptions 1–6, the inclusion
\[
\operatorname{supp}(\boldsymbol{\beta}_{t+1}^{(l)}) \subseteq \mathcal{S}
\]
holds for every $l = 1, \ldots, m$ with probability converging to one.
In addition, consider a sufficiently large constant $C > 0$ fulfilling
\begin{equation*}
\min_{k \in \mathcal{S}} |\beta_k^*| \geq C \left\| \left( \frac{1}{m} \sum_{l=1}^{m} \boldsymbol{\Sigma}_{l,\mathcal{S} \times \mathcal{S}} \right)^{-1} \right\|_{\mathrm{op}} \left( \sqrt{\frac{\log N}{N}} + \max_{1 \le l \le m} \{ a_{N,l} \} \sqrt{\frac{s \log N}{n}} + \gamma^T \right),
\end{equation*}
in which $\gamma \in (0,1)$ denotes the contraction factor introduced in Theorem~1(b).
Under such setup, the equality
\begin{equation*}
\operatorname{supp}(\boldsymbol{\beta}_{t+1}^{(l)}) = \mathcal{S}
\end{equation*}
is valid for all $l = 1, \ldots, m$ with probability tending to one.
This beta-min separation ensures nonzero coefficients remain separated away from zero, preventing erroneous shrinkage of true nonzero entries. The restriction remains consistent with the convergence bounds of $\boldsymbol{\beta}_{t+1}^{(l)}$ derived in Theorem~2.
\end{theorem}
This theorem establishes strong variable selection consistency for the proposed decentralized estimator. When combined with the linear convergence guarantee for the generalized consensus ADMM algorithm in Theorem 1 and the finite-sample error bound derived in Theorem 2, our theoretical framework fully characterizes the optimization convergence, statistical estimation precision, and support recovery performance of DeCRR. Taken together, these results verify that the proposed decentralized sparse convolution rank regression framework achieves reliable sparse estimation and consistent variable selection over decentralized networks even in the presence of heavy-tailed noise.

\section{Simulation studies}

\subsection{Design of the simulations}

A connected decentralized network containing $m$ nodes is built under the Erd\H{o}s--R\'enyi random graph framework \citep{bollobas2001random} with pairwise edge connection probability $p_c$. Local datasets residing at each node are produced following the linear regression model \eqref{eq:1}. In particular, covariates $\mathbf{x}_i$ constitute i.i.d. draws from multivariate Normal distribution $\mathcal{N}(\mathbf{0}, \boldsymbol{\Sigma})$, whose covariance matrix entries satisfy $\Sigma_{lk} = \sigma^2 \rho^{|l-k|}$ with positive constants $\sigma^2 > 0$ and $\rho > 0$. The true regression coefficient vector takes the form $\boldsymbol{\beta}^* = (3, 1.5, 0, 0, 2, 0, \ldots, 0)^{\top} \in \mathbb{R}^p$.

Numerical evaluations of the algorithm are conducted under three noise specifications:
\begin{enumerate}[(1)]
    \item Standard Normal distribution: $\epsilon \sim \mathcal{N}(0,1)$;
    \item Student's $t$-distribution with two degrees of freedom: $\epsilon \sim t(2)$;
    \item Cauchy distribution: $\epsilon \sim \text{Cauchy}(0,1)$.
\end{enumerate}
Experimental parameters are held constant at $m = 10$, $\sigma^2 = 0.3$, and $\rho = 0.1$ unless noted otherwise. All simulated samples are mutually independent and identically distributed.

The initial estimator at every node is obtained by solving $\ell_1$-penalized sparse convolution rank regression utilizing solely locally stored node-specific observations. We adopt a unified tuning parameter $\lambda_N$ shared across all nodes, which is selected via an aggregated Bayesian information criterion (BIC). This BIC-based tuning strategy originates from rank-based regression models in \citet{wang2009bic} and \citet{gu2020bic}. Given local candidate estimators $\{\hat{\boldsymbol{\beta}}^{(l)}\}_{l=1}^{m}$ across all $m$ nodes, the aggregated Bayesian information criterion is defined as follows:
\begin{equation}
\log\left(\frac{1}{m}\sum_{l=1}^{m} \frac{2}{n(n-1)} \sum_{\substack{i,j \in \mathcal{I}_l \\ j \neq i}} l_h\Big(y_i - y_j - (\mathbf{x}_i - \mathbf{x}_j)^{\top} \hat{\boldsymbol{\beta}}^{(l)}\Big)\right)
+ \frac{\log N}{N} \sum_{l=1}^{m} \big|\hat{\mathcal{S}}^{(l)}\big|,
\label{eq:bic}
\end{equation}
where a Normal kernel is employed to construct the local loss function, with the kernel form $K(u) = (1/\sqrt{2\pi})\exp(-u^2/2)$. Here, $\mathcal{S}=\operatorname{supp}(\boldsymbol{\beta}^*)$ stands for the true global support set, while $\hat{\mathcal{S}}^{(l)} = \operatorname{supp}(\hat{\boldsymbol{\beta}}^{(l)})$ signifies the estimated support set of local coefficients at node $l$, a quantity measuring the count of non-zero entries within $\hat{\boldsymbol{\beta}}^{(l)}$. In practical implementation, aggregating local loss and sparsity information to determine a common tuning parameter incurs only negligible communication overhead.

The kernel bandwidth is selected using the adaptive Silverman’s rule. Each node independently computes the local bandwidth as $h_l = c_h \cdot \operatorname{IQR}(y_l) \cdot 1.349^{-1} \cdot n^{-1/5}$, where $\operatorname{IQR}(y_l)$ denotes the interquartile range of local responses $y_l$. The global bandwidth is averaged over all nodes as $h = m^{-1}\sum_{l=1}^{m} h_l$, and $c_h$ is a constant scaling factor. The ADMM penalty parameter $\rho_l$ is set according to upper bounds of local Hessian matrices. For each node $l$, we take $\rho_l = (2\pi)^{-1/2} \cdot h_l^{-1} \cdot \lambda_{\max}\big(n^{-1}\mathbf{X}_l^{\top}\mathbf{X}_l\big) \cdot 1.1$; a mild safety buffer embodied by factor $1.1$ guarantees compliance with the theoretical step-size constraint outlined in Theorem 1. The global penalty parameter is further defined as $\tau = \gamma \cdot m^{-1}\sum_{l=1}^{m}\rho_l$, where the scaling coefficient $\gamma$ controls the overall penalty strength and is fixed to $\gamma = 0.1$ across all experimental trials.

Four quantitative evaluation criteria are utilized for comprehensive assessment of numerical performance for the approach developed in this work. The estimation error constitutes the first evaluation measure and takes the form $\bigl( \frac{1}{m} \sum_{l=1}^m \| \hat{\boldsymbol{\beta}}^{(l)} - \boldsymbol{\beta}^* \|_2^2 \bigr)^{1/2}$. The second measure corresponds to network-averaged sparsity precision. For each node $l$, this quantity describes the share of correctly detected nonzero coefficients among all estimated nonzero entries, expressed as $\text{Precision}^{(l)} = |\hat{\mathcal{S}}^{(l)} \cap \mathcal{S}| / |\hat{\mathcal{S}}^{(l)}|$. Network-averaged sparsity recall serves as the third criterion, which characterizes the fraction of genuine nonzero coefficients recovered locally at node $l$ according to $\text{Recall}^{(l)} = |\hat{\mathcal{S}}^{(l)} \cap \mathcal{S}| / |\mathcal{S}|$. The fourth evaluation measure is the network-averaged $F_1$-score, representing the harmonic mean of node-wise precision and recall: $\text{F1-score}^{(l)} = 2 \cdot (\text{Precision}^{(l)} \cdot \text{Recall}^{(l)}) / (\text{Precision}^{(l)} + \text{Recall}^{(l)})$. Values of the $F_1$-score lie within $[0,1]$, and greater magnitudes reflect superior support recovery behavior. Such criterion is commonly leveraged throughout high-dimensional sparse regression studies to quantify variable selection fidelity.

All experiments are repeated over $100$ independent replications to ensure stable and reliable numerical results.

\subsection{Effect of iterations}
Influence of ADMM iteration counts upon estimation convergence and sparse variable selection performance is explored within this subsection under Normal, $t(2)$, and Cauchy noise, while simulation configurations remain fixed with $n=200$ and $p=100$.

\begin{figure}[htbp]
\centering
\includegraphics[width=0.8\textwidth]{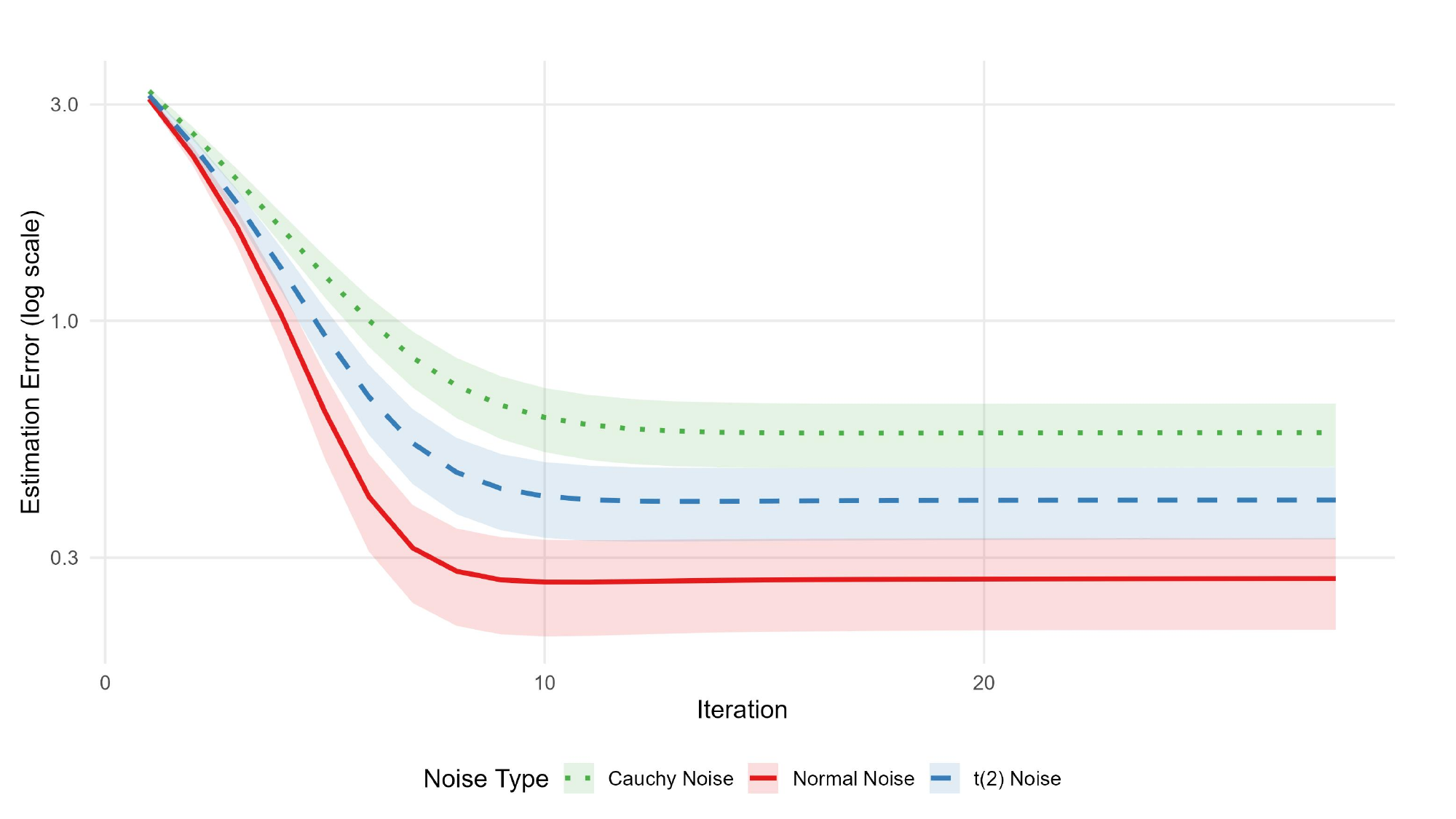}
\caption{Log-scale estimation error versus iterations under Normal, $t(2)$, and Cauchy noise. Solid lines denote different noise scenarios, and shaded bands indicate standard deviations across replications.}
\label{fig:conv_noise}
\end{figure}

Log-scale estimation error trajectories against iteration count are visualized in Figure \ref{fig:conv_noise}, with shaded regions representing the standard deviation across independent replications. For all three noise settings, the estimation error decreases rapidly in early iterations, which empirically validates the linear convergence property derived in Theorem 1. All curves eventually converge to a stable asymptotic error floor, yielding a clear performance ranking that demonstrates the inherent robustness of the convolution rank loss against heavy-tailed noise. As the iteration count increases, the optimization residual decays exponentially and gradually becomes negligible. Accordingly, the steady-state estimation bias primarily arises from kernel smoothing approximations and heterogeneity in local initial estimates. Notably, faster convergence is observed under heavy-tailed noise, and stable convergence is consistently achieved within the predefined iteration budget across all scenarios.

\begin{figure}[htbp]
\centering
\includegraphics[width=0.65\textwidth]{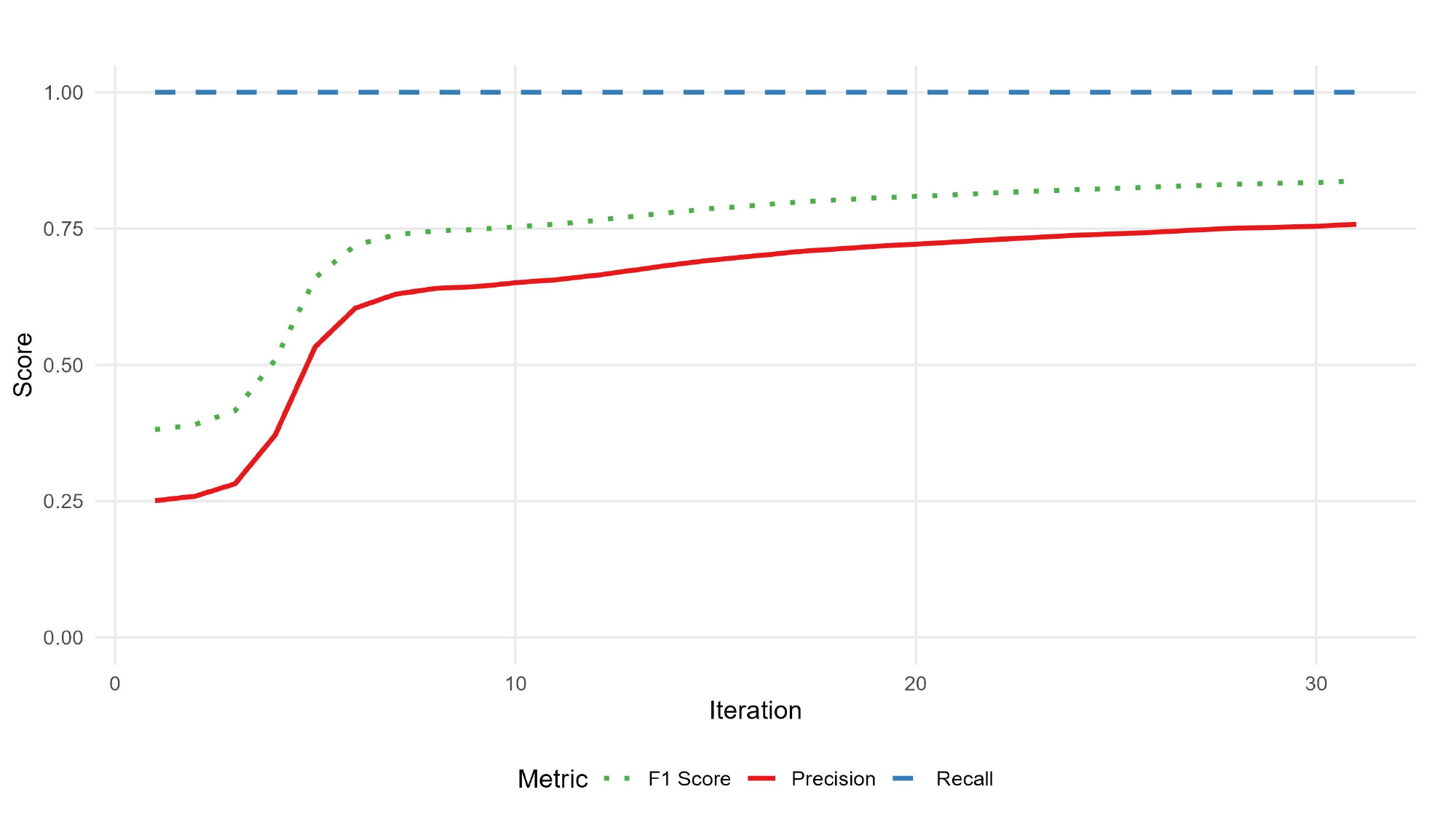}
\vspace{4pt}
\includegraphics[width=0.65\textwidth]{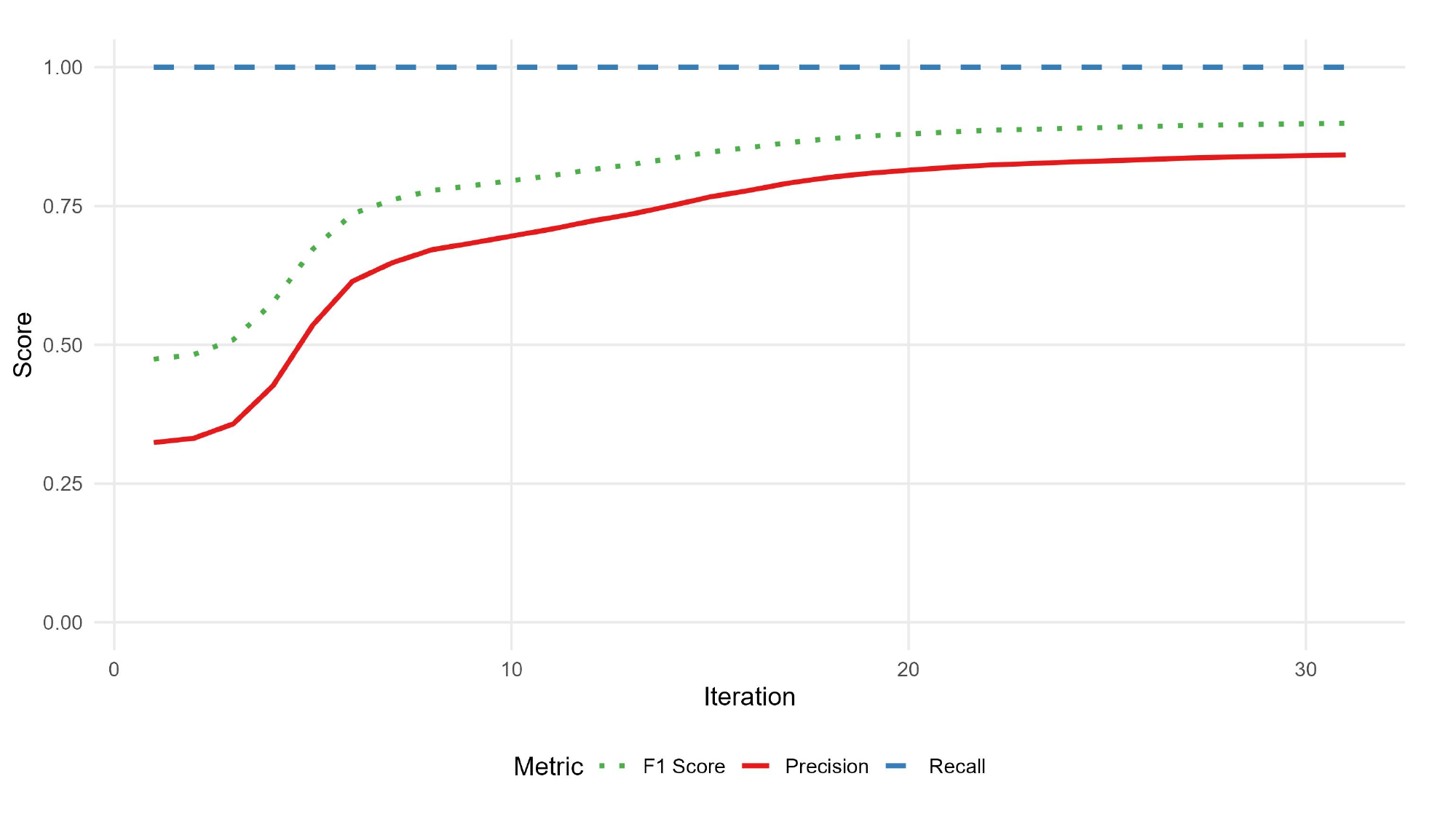}
\vspace{4pt}
\includegraphics[width=0.65\textwidth]{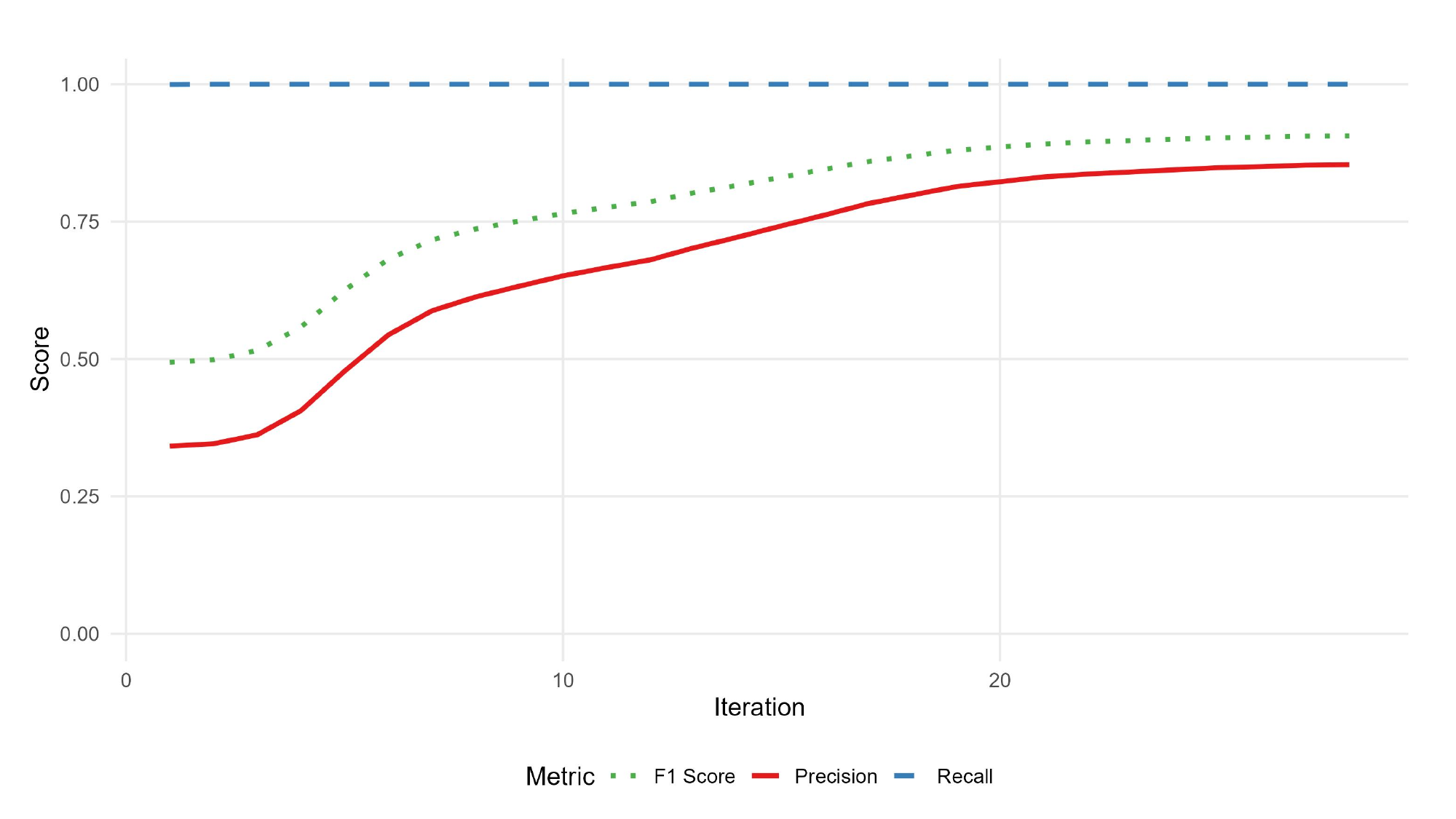}
\caption{Recall, Precision, and F1-score versus iterations under Normal, $t(2)$, and Cauchy noise. Shaded bands represent standard deviations over replications.}
\label{fig:var_sel_all}
\end{figure}

Iteration-varying Recall, Precision, and F1-score under the three noise distributions are further illustrated in Figure \ref{fig:var_sel_all}.
The recall remains identically one throughout the entire iterative process, implying full recovery of all true predictive variables at any iteration. In comparison, Precision and F1-score start at low levels and improve substantially in early iterations, as decentralized consensus updates effectively mitigate false-positive variable selections. All evaluation metrics stabilize after adequate iterations and maintain reliable sparsity identification under both light-tailed and heavy-tailed noise, which verifies the stable and competitive variable selection performance of the proposed method across diverse noise environments.

\subsection{Effect of the CRR loss under different noises}

Robustness of DeCRR against heavy-tailed noise and its competitive estimation efficiency under light-tailed noise are jointly assessed by accounting for the noise distributions outlined above. A decentralized $\ell_1$-penalized least squares estimator (deLR) is further established, which adopts the generalized consensus ADMM derived in Section 2.4 for the $\ell_1$-regularized least squares objective function. The maximum iteration limit for both schemes is set to 1000 to guarantee fair comparisons between deLR and DeCRR.

Estimation errors, precision and recall are summarized in Table \ref{tab:1}. Under Normal noise, DeCRR and deLR deliver comparable performance. In contrast, DeCRR substantially outperforms deLR when the errors follow the $t(2)$ or Cauchy distribution. Accordingly, we exclude the decentralized least squares estimator from all subsequent comparative experiments.

\begin{table}[H]
\centering
\caption{Estimation error, precision, and recall obtained from deLR and DeCRR estimators under Normal, Student's $t(2)$, and Cauchy noise distributions, with local sample sizes and dimensions $(n,p)$ varied over the set $\{(100,100),(200,100),(200,200)\}$.}
\label{tab:1}
\begin{tabular}{ccccccc}
\toprule
\multirow{2}{*}{$(n,p)$} & \multicolumn{3}{c}{deLR} & \multicolumn{3}{c}{DeCRR} \\
\cmidrule{2-4} \cmidrule{5-7}
 & Est.error & Precision & Recall & Est.error & Precision & Recall \\
\midrule
\multicolumn{7}{c}{Normal Noise} \\
(150,100) & 0.1573 & 0.9294 & 1.0000 & 0.2780 & 0.9277 & 1.0000 \\
(200,100) & 0.1507 & 0.9110 & 1.0000 & 0.2523 & 0.9356 & 1.0000 \\
(200,200) & 0.1460 & 0.9504 & 1.0000 & 0.2541 & 0.9372 & 1.0000 \\
\midrule
\multicolumn{7}{c}{$t(2)$ Noise} \\
(150,100) & 0.3365 & 0.1288 & 1.0000 & 0.3813 & 0.9477 & 1.0000 \\
(200,100) & 0.3210 & 0.1711 & 1.0000 & 0.3221 & 0.9476 & 1.0000 \\
(200,200) & 0.3311 & 0.1089 & 1.0000 & 0.3108 & 0.9530 & 1.0000 \\
\midrule
\multicolumn{7}{c}{Cauchy Noise} \\
(150,100) & 408.4529 & 0.0318 & 0.9730 & 0.4297 & 0.9590 & 1.0000 \\
(200,100) & 61.6911 & 0.0323 & 0.9883 & 0.4001 & 0.9539 & 1.0000 \\
(200,200) & 89.2496 & 0.0164 & 0.9966 & 0.3983 & 0.9361 & 1.0000 \\
\bottomrule
\end{tabular}
\end{table}

\subsection{Effect of heterogeneity}

This section evaluates the performance of DeCRR under heterogeneous data configurations where different nodes follow distinct data-generating distributions. We consider two experimental setups:
(1) Node-specific covariates are generated from heterogeneous AR(1) structures with uniform Cauchy noise. Each node independently draws $\sigma^2 \in \{1,3\}$ and autocorrelation coefficient $\rho \in \{0.1,0.3\}$, and the corresponding AR(1) covariance matrix satisfies $\Sigma_{ij} = \sigma^2 \rho^{|i-j|}$.
(2) Nodes are assigned heterogeneous noise randomly sampled from the standard Normal, heavy-tailed $t(2)$, and extremely heavy-tailed Cauchy distributions, whereas covariates across all nodes share the same AR(1) covariance structure with $\sigma^2=1$ and $\rho=0.1$.

We compare DeCRR with four competing estimators, defined as follows:

(1) Pooled CRR: a centralized estimator trained on the fully aggregated dataset,
\[
\hat{\boldsymbol{\beta}}_{\text{pool}} = \arg\min_{\boldsymbol{\beta} \in \mathbb{R}^p} \left\{ \frac{1}{N(N-1)} \sum_{i=1}^{N}\sum_{j \neq i} l_h\big(y_i - y_j - (\mathbf{x}_i - \mathbf{x}_j)^{\top}\boldsymbol{\beta}\big) + \lambda_N \|\boldsymbol{\beta}\|_1 \right\},
\]
which serves as the performance upper bound. All metrics are evaluated directly on this single global estimate.

(2) Local CRR: node-wise estimator trained using only local samples without inter-node communication,
\[
\hat{\boldsymbol{\beta}}_{\text{local}}^{(l)} = \arg\min_{\boldsymbol{\beta} \in \mathbb{R}^p} \left\{ \frac{1}{n(n-1)} \sum_{\substack{i,j \in \mathcal{I}_l \\ j \neq i}} l_h\big(y_i - y_j - (\mathbf{x}_i - \mathbf{x}_j)^{\top}\boldsymbol{\beta}\big) + \lambda_n \|\boldsymbol{\beta}\|_1 \right\}, \quad l = 1,\ldots,m.
\]

Metrics are computed separately for each node, and we report the average across all $m$ nodes.

(3) Avg CRR: constructed by averaging all independent local CRR estimates to yield a unified global estimator,
\[
\hat{\boldsymbol{\beta}}_{\text{avg}} = \frac{1}{m}\sum_{l=1}^m \hat{\boldsymbol{\beta}}_{\text{local}}^{(l)},
\]

Evaluation metrics are calculated based on the aggregated global estimate $\hat{\boldsymbol{\beta}}_{\text{avg}}$. Note that this simple averaging destroys sparsity by producing a dense vector from local estimates with distinct supports, thus cannot guarantee consistent variable selection in high-dimensional settings.

(4) D-subGD: decentralized subgradient descent, a distributed subgradient optimization algorithm proposed by \citet{wangli2018}.

Numerical results including estimation error and $F_1$-score are summarized in Tables~\ref{tab:2} and~\ref{tab:3}. As shown in the tables, the proposed DeCRR achieves performance comparable to pooled CRR under both heterogeneous settings and substantially outperforms the other three competing methods. In particular, the local estimator relies exclusively on data from an individual node and cannot leverage global information, thereby yielding relatively poor performance. The simple averaging strategy fails to compensate for distributional discrepancies across nodes and leads to unsatisfactory variable selection. Although its estimation error is similar to that of the local estimator, it remains limited in recovering the underlying sparse structure. D-subGD delivers moderate variable selection performance but exhibits markedly inferior estimation accuracy, with considerably larger estimation error than DeCRR. This observation aligns with the known theoretical property that subgradient methods typically converge slowly and are sensitive to step-size tuning. By contrast, DeCRR achieves accurate parameter estimation and reliable variable selection simultaneously in distributed heterogeneous environments, while maintaining strong robustness against heterogeneous data distributions.

\begin{table}[H]
\centering
\caption{Estimation errors yielded by Pooled CRR, Local CRR, Avg CRR, D-subGD, and DeCRR under different data heterogeneity configurations.}
\label{tab:2}
\begin{tabular}{cccccc}
\toprule
$(n,p)$ & Pooled CRR & Local CRR & Avg CRR & D-subGD & DeCRR \\
\midrule
\multicolumn{6}{c}{Covariate Heterogeneity} \\
(150,100) & 0.1814 & 0.6273 & 0.5618 & 1.0236 & 0.3559 \\
(100,200) & 0.1608 & 0.5722 & 0.5300 & 0.9837 & 0.3521 \\
(200,200) & 0.1606 & 0.5945 & 0.5476 & 1.0083 & 0.2946 \\
\midrule
\multicolumn{6}{c}{Noise Heterogeneity} \\
(150,100) & 0.2171 & 0.6698 & 0.6388 & 1.0438 & 0.3719 \\
(100,200) & 0.2121 & 0.6062 & 0.5808 & 0.9720 & 0.3391 \\
(200,200) & 0.2108 & 0.6520 & 0.6282 & 1.0296 & 0.2901 \\
\bottomrule
\end{tabular}
\end{table}

\begin{table}[H]
\centering
\caption{$F_1$-scores achieved by Pooled CRR, Local CRR, Avg CRR, D-subGD, and DeCRR under different data heterogeneity scenarios.}
\label{tab:3}
\begin{tabular}{cccccc}
\toprule
$(n,p)$ & Pooled CRR & Local CRR & Avg CRR & D-subGD & DeCRR \\
\midrule
\multicolumn{6}{c}{Covariate Heterogeneity} \\
(150,100) & 0.9657 & 0.6725 & 0.1489 & 0.2796 & 0.9678 \\
(100,200) & 0.9557 & 0.7805 & 0.2298 & 0.4149 & 0.9724 \\
(200,200) & 0.9461 & 0.6756 & 0.1297 & 0.2690 & 0.9904 \\
\midrule
\multicolumn{6}{c}{Noise Heterogeneity} \\
(150,100) & 0.9986 & 0.9369 & 0.5883 & 0.8093 & 0.9830 \\
(100,200) & 0.9986 & 0.9801 & 0.8307 & 0.9472 & 0.9836 \\
(200,200) & 0.9943 & 0.9626 & 0.7127 & 0.8861 & 0.9914 \\
\bottomrule
\end{tabular}
\end{table}

\subsection{Effect of network topology}

Impacts of network topology upon DeCRR performance are analyzed within this section, with attention directed toward two critical factors: node count and network sparsity.

The overall sample size is fixed at $N = 4200$, with three node configurations $m = 5, 10,$ and $20$ adopted for evaluation. To reduce fluctuations caused by network randomness, we employ fully connected topologies. Comparisons between DeCRR and three alternative decentralized schemes are summarized within Table~\ref{tab:4}. In terms of estimation error, the proposed DeCRR outperforms all competitors under nearly all experimental setups. For support recovery, DeCRR sustains favorable performance with only mild degradation as the number of nodes grows or under heavy-tailed noise. These comparisons verify that DeCRR attains high estimation accuracy and stable variable selection simultaneously under Normal, $t(2)$, and Cauchy noise, and highlight its distinct advantages for robust sparse regression over decentralized networks.

This analysis further explores network sparsity effects by varying the edge connection probability across $0.3$, $0.5$, and $0.8$. The network configuration is fixed with $m = 10$ nodes, a local sample size of $n = 200$, and a feature dimension of $p = 50$. Table~\ref{tab:5} presents the estimation error, precision, and recall corresponding to different connection probability settings. DeCRR yields comparable performance across all sparsity levels. This phenomenon arises because network sparsity only enters the convergence constant $\gamma$. As established in Theorem 1, the error term governed by $\gamma$ decays linearly and becomes asymptotically negligible relative to the remaining statistical error components.

\begin{table}[H]
\centering
\caption{Estimation errors and $F_1$-scores attained by Local CRR, Avg CRR, D-subGD, and DeCRR under Normal, Student's $t(2)$, and Cauchy noise distributions, with the node number $m$ varied over $\{5, 10, 20\}$.}
\label{tab:4}
\small
\setlength{\tabcolsep}{4pt}
\begin{tabular}{ccccccccc}
\toprule
\multirow{2}{*}{$m$} & \multicolumn{2}{c}{Local CRR} & \multicolumn{2}{c}{Avg  CRR} & \multicolumn{2}{c}{D-subGD} & \multicolumn{2}{c}{DeCRR} \\
\cmidrule{2-3} \cmidrule{4-5} \cmidrule{6-7} \cmidrule{8-9}
 & Est. err & F1 & Est. err & F1 & Est. err & F1 & Est. err & F1 \\
\midrule
\multicolumn{9}{c}{Normal Noise} \\
5 & 0.1952 & 0.9877 & 0.1892 & 0.9400 & 0.4217 & 1.0000 & 0.1566 & 0.9994 \\
10 & 0.2742 & 0.9928 & 0.2644 & 0.9312 & 0.5719 & 1.0000 & 0.1601 & 0.9784 \\
20 & 0.3864 & 0.9852 & 0.3706 & 0.7686 & 0.7628 & 0.9800 & 0.3048 & 0.9848 \\
\midrule
\multicolumn{9}{c}{Student's $t(2)$ Noise} \\
5 & 0.2740 & 0.9764 & 0.2653 & 0.8935 & 0.4872 & 1.0000 & 0.2166 & 1.0000 \\
10 & 0.3927 & 0.9898 & 0.3783 & 0.9050 & 0.6711 & 1.0000 & 0.2565 & 0.9756 \\
20 & 0.5662 & 0.9835 & 0.5422 & 0.7404 & 0.9186 & 0.9671 & 0.4649 & 0.9946 \\
\midrule
\multicolumn{9}{c}{Cauchy Noise} \\
5 & 0.3464 & 0.9541 & 0.3340 & 0.8040 & 0.5381 & 1.0000 & 0.2893 & 1.0000 \\
10 & 0.5248 & 0.9672 & 0.5040 & 0.7408 & 0.7849 & 1.0000 & 0.3709 & 0.9758 \\
20 & 0.8178 & 0.9763 & 0.7822 & 0.6677 & 1.1494 & 0.9525 & 0.6476 & 0.9906 \\
\bottomrule
\end{tabular}
\end{table}

\begin{table}[H]
\centering
\caption{Estimation error, precision, and recall obtained from the DeCRR estimator under Normal, Student's $t(2)$, and Cauchy noise distributions, where network connection probability $p_c$ is chosen from $\{0.3, 0.5, 0.8\}$.}
\label{tab:5}
\small
\setlength{\tabcolsep}{4pt}
\begin{tabular}{cccccccccc}
\toprule
\multirow{2}{*}{$p_c$} & \multicolumn{3}{c}{Normal} & \multicolumn{3}{c}{Student's $t(2)$} & \multicolumn{3}{c}{Cauchy} \\
\cmidrule{2-4} \cmidrule{5-7} \cmidrule{8-10}
 & Err & Prec & Rec & Err & Prec & Rec & Err & Prec & Rec \\
\midrule
0.3 & 0.2571 & 0.9657 & 1.0000 & 0.3540 & 0.9644 & 1.0000 & 0.4419 & 0.9618 & 1.0000 \\
0.5 & 0.2495 & 0.9697 & 1.0000 & 0.2969 & 0.9592 & 1.0000 & 0.3845 & 0.9552 & 1.0000 \\
0.8 & 0.2445 & 0.9525 & 1.0000 & 0.3812 & 0.9524 & 1.0000 & 0.5258 & 0.9475 & 1.0000 \\
\bottomrule
\end{tabular}
\end{table}

\subsection{Effect of dimension}

This section further investigates how feature dimensionality affects the numerical performance of the DeCRR method. With the local sample size fixed at $n = 100$, the feature dimension is varied across $p \in \{100, 200, 300, 400, 500\}$. The highest dimensional setting adopted in this work aligns with the experimental designs presented in \citet{chen2023distcqr} and \citet{chen2024byz}, both of which employ comparable high-dimensional evaluation protocols. Notably, our setup imposes a more challenging scenario, as our local sample size $n$ is substantially smaller. This study compares DeCRR with the four competing algorithms outlined in Section 4.4. Table~\ref{tab:6} and Table~\ref{tab:7} summarize the estimation errors and $F_1$-scores across different dimensional settings, respectively.

Numerical results demonstrate that DeCRR is robust to increasing dimensionality, consistently maintaining low estimation error and high $F_1$-score even in the small-sample and high-dimensional regime. In contrast, the competing Local CRR, Avg CRR, and D-subGD methods suffer substantial performance degradation as $p$ grows, further verifying the superior dimensional robustness of the proposed DeCRR estimator.

\begin{table}[H]
\centering
\caption{Estimation errors attained by Local CRR, Avg CRR, D-subGD, and DeCRR under Cauchy noise with varying feature dimension $p$.}
\label{tab:6}
\begin{tabular}{cccccc}
\toprule
$p$ & Pooled CRR & Local CRR & Avg  CRR & D-subGD & DeCRR \\
\midrule
100 & 0.3141 & 1.0502 & 0.9780 & 1.3675 & 0.8822 \\
200 & 0.3416 & 1.0950 & 0.9949 & 1.3854 & 0.8969 \\
300 & 0.3774 & 1.1227 & 1.0129 & 1.4022 & 0.9074 \\
400 & 0.4068 & 1.1733 & 1.0504 & 1.4407 & 0.9223 \\
500 & 0.3886 & 1.1741 & 1.0382 & 1.4280 & 0.8976 \\
\bottomrule
\end{tabular}
\end{table}

\begin{table}[H]
\centering
\caption{$F_1$-scores of Local CRR, Avg CRR, D-subGD, and DeCRR under Cauchy noise, evaluated with varying dimension $p$.}
\label{tab:7}
\begin{tabular}{cccccc}
\toprule
$p$ & Pooled CRR & Local CRR & Avg  CRR & D-subGD & DeCRR \\
\midrule
100 & 0.9614 & 0.7411 & 0.2266 & 0.3076 & 0.9941 \\
200 & 0.9504 & 0.5844 & 0.1272 & 0.1783 & 0.9905 \\
300 & 0.9600 & 0.4907 & 0.0911 & 0.1266 & 0.9865 \\
400 & 0.9557 & 0.4177 & 0.0693 & 0.0951 & 0.9760 \\
500 & 0.9575 & 0.3658 & 0.0569 & 0.0779 & 0.9723 \\
\bottomrule
\end{tabular}
\end{table}

\subsection{Sensitivity analysis of bandwidth and penalty parameters}

We carry out a sensitivity analysis to investigate how the bandwidth coefficient $c_h$ and step-size coefficient $\eta$ within the generalized consensus ADMM affect the estimation performance of DeCRR. Specifically, we consider $c_h \in \{0.5, 1.06\}$ and $\eta \in \{0.1, 0.5, 1.0, 2.0\}$, and conduct experiments under three sample dimension configurations $(n,p) \in \{(150,100),\allowbreak (200,100),\allowbreak (200,200)\}$. Tables~\ref{tab:8} and~\ref{tab:9} collect numerical outcomes associated with different selections of $c_h$ and $\eta$, respectively. As shown in Table~\ref{tab:8}, the model achieves its best performance with a moderately small $c_h$; as $c_h$ increases, the estimation error grows markedly and the $F_1$-score degrades considerably. Numerical outcomes for different selections of $\eta$ are compiled in Table~\ref{tab:9}, while the $F_1$-score remains consistently high across the full tested ranges of both parameters.

\begin{table}[H]
\centering
\caption{Estimation error and $F_1$-score of the DeCRR estimator evaluated with varying $\eta$ under Cauchy noise.}
\label{tab:8}
\begin{tabular}{ccccccc}
\toprule
\multirow{2}{*}{$\eta$} & \multicolumn{3}{c}{Est. error} & \multicolumn{3}{c}{F1-score} \\
\cmidrule{2-4} \cmidrule{5-7}
 & $(150,100)$ & $(200,100)$ & $(200,200)$ & $(150,100)$ & $(200,100)$ & $(200,200)$ \\
\midrule
0.1 & 0.6396 & 0.5570 & 0.5456 & 0.9627 & 0.9651 & 0.9632 \\
0.5 & 0.7395 & 0.6083 & 0.7850 & 0.9533 & 0.9459 & 0.9632 \\
1.0 & 0.7914 & 0.6241 & 0.6632 & 0.9629 & 0.9315 & 0.9391 \\
2.0 & 0.6143 & 0.5476 & 0.4942 & 0.9665 & 0.9737 & 0.9712 \\
\bottomrule
\end{tabular}
\end{table}

\begin{table}[H]
\centering
\caption{Estimation error and $F_1$-score of the DeCRR estimator evaluated with varying $c_h$ under Cauchy noise.}
\label{tab:9}
\begin{tabular}{ccccccc}
\toprule
\multirow{2}{*}{$c_h$} & \multicolumn{3}{c}{Est. error} & \multicolumn{3}{c}{F1-score} \\
\cmidrule{2-4} \cmidrule{5-7}
 & $(150,100)$ & $(200,100)$ & $(200,200)$ & $(150,100)$ & $(200,100)$ & $(200,200)$ \\
\midrule
0.5 & 0.4647 & 0.4040 & 0.3989 & 0.9791 & 0.9745 & 0.9653 \\ 					
1.06 & 0.6412 & 0.5789 & 0.5493 & 0.9607 & 0.9667 & 0.9667 \\ 					
\bottomrule
\end{tabular}
\end{table}

Collectively, these experimental outcomes demonstrate that variable selection performance remains robust against fluctuations in both the bandwidth coefficient and the step-size coefficient. The sensitivity analysis further confirms that DeCRR maintains stable and favorable performance over reasonable parameter ranges. Within these ranges, the estimation error remains low and the $F_1$-score stays persistently high. This feature highlights the reliable robustness of the proposed method under appropriate parameter configurations.

\section{Communities and Crime data analysis}
Further validation of DeCRR's practical applicability is conducted on the real-world Communities and Crime dataset retrieved from the UCI Machine Learning Repository. Constructed and released by \citet{redmond2009crime}, this dataset is adopted herein to identify key demographic and socioeconomic factors associated with community-level crime rates and characterize their underlying linear relationships.

This dataset consolidates three official data streams: socioeconomic statistics from the 1990 U.S. Census, law enforcement records from the 1990 Law Enforcement Management and Administrative Statistics Survey, and crime reports from the 1995 FBI Uniform Crime Reporting Program.
The original dataset comprises 147 attributes collected over 2,215 communities spanning 49 U.S. states. The per-capita violent crime rate per 100,000 residents (ViolentCrimesPerPop) serves as the response variable in the regression analysis. Raw records undergo preprocessing via removal of incomplete samples and standardization of all feature variables, which produces a cleaned dataset consisting of 90 predictors and 1,901 community observations. Subsequently, the processed data are randomly split into an 80\% training subset and a 20\% test subset.

\begin{figure}[H]
\centering
\includegraphics[width=0.6\textwidth]{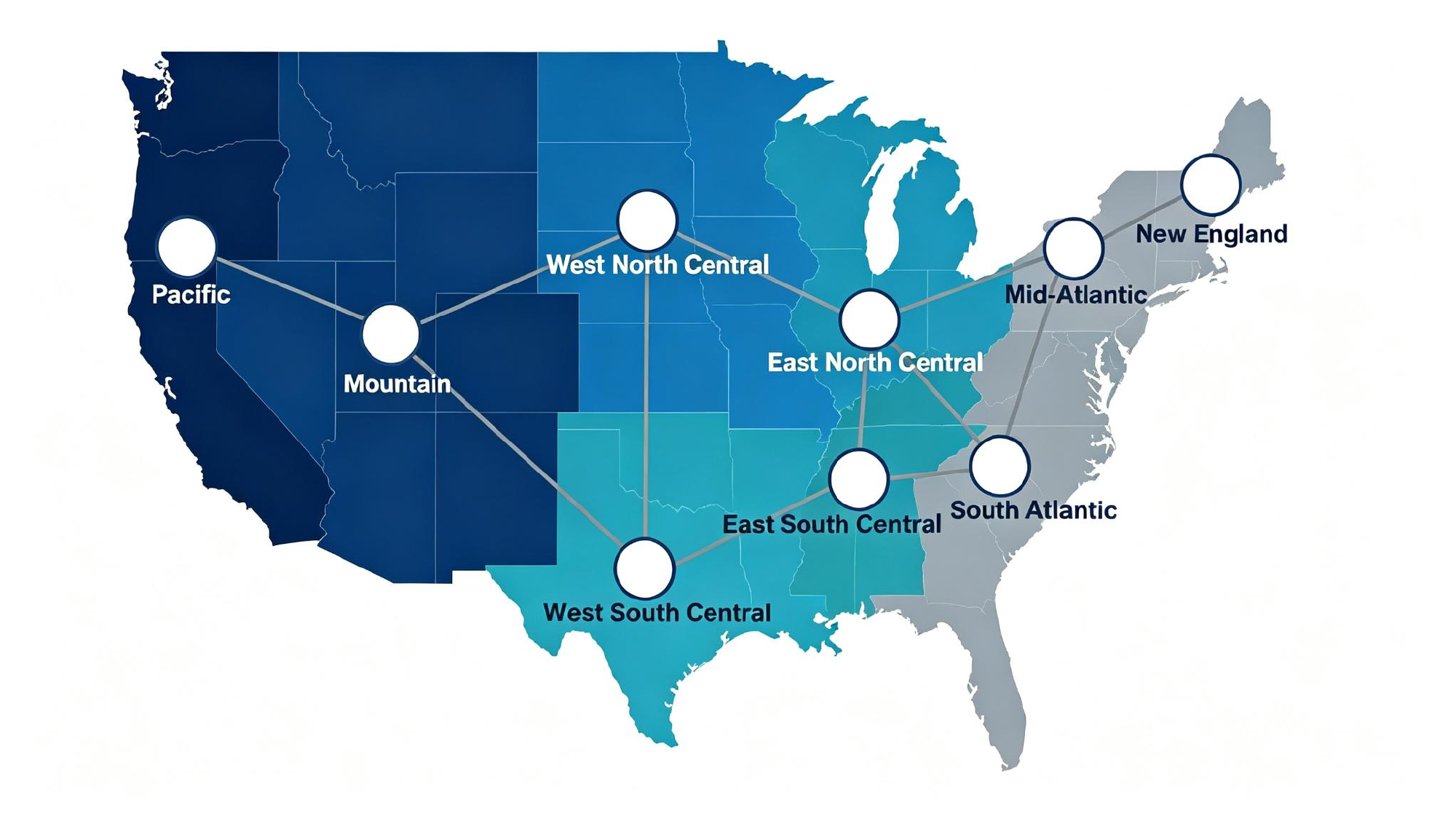}
\caption{Network topology of the nine geographic division units.}
\label{fig:network}
\end{figure}

Following the preprocessing protocol proposed by \citet{yang2019}, we assign each community to one of nine geographical divisions defined by the U.S. Census Bureau. Each division corresponds to a local data node within our decentralized framework and stores its community-specific samples. Under this distributed setting, each node only exchanges information with its geographic neighbors, forming the network topology visualized in Figure~\ref{fig:network}.

Within each geographical division, 80\% of local samples are randomly assigned to train DeCRR, Avg CRR, and D-subGD, whereas the remaining 20\% local samples serve to calculate the root mean squared error (RMSE) and mean absolute error (MAE).
Table~\ref{tab:10} aggregates averaged test metrics across all nine divisions.
Empirical findings on the real dataset reveal that the proposed DeCRR achieves considerably better performance than the two competing benchmarks under both evaluation criteria.

\begin{table}[H]
\centering
\caption{RMSE and MAE performance of the Avg CRR, D-subGD, and DeCRR estimators on the Communities and Crime dataset.
}
\label{tab:10}
\begin{tabular}{cccc}
\toprule
Metric & Avg CRR & D-subGD & DeCRR \\
\midrule
RMSE & 0.8136 & 1.8030 & 0.7850 \\
MAE & 0.5854 & 1.3684 & 0.5662 \\
\bottomrule
\end{tabular}
\end{table}

\section{Conclusion}
This work proposes a novel decentralized collaborative algorithm for penalized convolution rank regression. Built upon the generalized consensus ADMM framework, the proposed method solves the optimization problem of convolution rank regression in a fully decentralized peer-to-peer manner. Both theoretical derivations and numerical simulations validate the effectiveness and practical superiority of our algorithm. Notably, the theoretical guarantees derived in this study apply exclusively to linear regression settings. Extending the proposed framework to broader scenarios, including generalized linear models and semiparametric regression, constitutes a promising direction for future research.

\bibliographystyle{plainnat}
\bibliography{ref}

\end{document}